\documentclass{article}
\usepackage{amssymb}
\usepackage{graphicx}
\usepackage{cite}
\begin{document}

\title{\bf Radiation of Charged Black Holes and Modified Dispersion Relation}
\author{A. D. Kamali and P. Pedram\thanks{Email: p.pedram@srbiau.ac.ir (Corresponding Author)}\
\\{\small Department of Physics, Science and Research Branch, Islamic Azad University, Tehran, Iran}}

\maketitle \baselineskip 24pt

\begin{abstract}
We investigate the effects of a modified dispersion relation
proposed by Majhi and Vagenas on the Reissner-Nordstr\"{o}m black
hole thermodynamics in a universe with large extra dimensions. It is
shown that entropy, temperature and heat capacity receive new
corrections and charged black holes in this framework have less
degrees of freedom and decay faster compared to black holes in the
Hawking picture. We also study the emission rate of black hole and
compare our results with other quantum gravity approaches. In this
regard, the existence of the logarithmic prefactor and the relation
between dimensions and charge are discussed. This procedure is not
only valid for a single horizon spacetime but it is also valid for
the spacetimes with inner and outer horizons.

\vspace{.5cm} {\it Keywords}: Black hole physics, Modified
dispersion relations, Reissner-Nordstr\"{o}m black hole
\end{abstract}
\maketitle

\section{Introduction}
A common feature of all promising candidates for quantum gravity
such as string theory \cite{Veneziano86}, loop  quantum gravity
\cite{Garay95}, noncommutative geometry \cite{Capozziello00},  and
black hole physics is the existence of a minimum observable length
\cite{Meissner04,Adler01}. This minimum measurable length gives rise
to the modification of Heisenberg uncertainty principle, nowadays
known as Generalized Uncertainty Principle (GUP) \cite{Kempf95}.  On
the other hand, in the context of doubly special relativity (DSR)
theories \cite{Camelia02,Magueijo05}, in order to preserve the
velocity of light and the planck energy as two invariant quantities,
the existence of a maximal momentum is essentially required. So, it
provides several novel and interesting features, some of which are
studied in \cite{Das08,Ali09,Das10,Etemadi12,Pedram11}. Also, DSR
motivates the modified dispersion relation
(MDR)\cite{Camelia01,Magueijo03}, by the fact that all approaches to
quantum gravity suggest that standard energy-momentum dispersion
relation should be modified near the Planck energy. This deformation
of the energy-momentum relation has been also suggested by the
discreteness of the spacetime \cite{Hooft96}.

Recently, much attention has been devoted to resolving the quantum
corrections to the black hole entropy. Identifying the microstates
is one of the main problems in studying the entropy of black holes.
Leading candidate theories of quantum gravity such as string theory
and loop quantum gravity predict the following entropy 
\cite{Kaul00,Medved04,Domagala04,Akbar04}
\begin{equation}
\label{e2} S= \frac{A}{4l_p^2} + c_0~\ln
\bigg(\frac{A}{4l_p^2}\bigg) + \sum_{n=1}^\infty c_n
\bigg(\frac{A}{4l_p^2}\bigg)^{-n} + \mbox{const.},
\end{equation}
where the coefficients $c_n$ can be regarded as model dependent
parameters. Black holes are suitable examples of an extreme quantum
gravity regime. Thus, study their thermodynamical behavior using MDR
and comparing the results with other approaches may increase our
understanding of their properties and structures. Indeed, the exact
form of MDR could essentially lead us to a deeper understanding of
the ultimate quantum gravity proposal.

Studying the thermodynamics of black holes in the presence of extra
dimensions which is an interesting issue is the subject of this work
\cite{Sefiedgar10,Sefiedgar11,Sefiedgar12}. Large extra dimension
models (LED) offer exciting ways to solve the hierarchy problem and
to study low scale quantum gravity effects
\cite{Arkani99,Randall99,Dvali00}. From a theoretical point of view,
one can expect that the properties of black holes may play an
important role in understanding the nature of gravity in higher
dimensions. The black hole and brane production in the LHC is also
studied in Ref.~\cite{Meade08}. Thus, it is important to investigate
the effects of extra dimensions on the various properties of black
holes.

In this paper, we intend to extend above analysis to the
Reissner-Nordstr\"{o}m (RN) black holes and we choose a specific
form of MDR proposed by Majhi and Vagenas, in which both energy and
momentum of particles are bounded. The organization of this work is
as follows: In Sec.~\ref{sec2}, we introduce briefly the modified
dispersion relation (MDR*) which admits minimal length and maximal
momentum. In Sec.~\ref{sec3}, we investigate a charged black hole
thermodynamics in universes with large extra dimensions. In
Secs.~\ref{sec4}-\ref{sec7}, we obtain entropy, temperature and heat
capacity of charged black hole in the presence of (MDR*). Also, we
find new corrections in the emission rate of charged black holes.
Finally, we present our conclusions in Sec.~\ref{sec8}.

\section{The modified dispersion relation (MDR*)}\label{sec2}
The idea of modified energy-momentum dispersion relation which is
known as MDR is popular among those are interesting in effective
approach to quantum gravity problems. The modified dispersion
relations are usually in the form \cite{Camelia06,Majumder}
\begin{equation}
p^2 \simeq E^2 - \mu^2 + \alpha_1 l_p E^3 + \alpha_2 l_p^2 E^4 +
\dots,
\end{equation}
where $\mu$ is the mass parameter and corresponds to the
rest energy of the particle. These  modified dispersion
relations have been used to calculate the black hole entropy (see
Ref.~\cite{Camelia06} for a brief discussion).

Recently, a new form of MDR  \cite{Vagenas13} (so called MDR*), has
been introduced which implies a minimum measurable  length and a
maximum measurable momentum.
\begin{eqnarray}\label{momentum}
p^0 = k^0,\qquad p^{i} = k^{i}(1-\alpha k + 2\alpha^2k^2),
\end{eqnarray}
where $k=|\bf{k}|$, $p^a$  is the momentum at high energies, and
$k^a$ is the momentum at low energies which satisfies the ordinary
dispersion relation. The gravitational background metric can be
considered as ($g_{0i}=0$)
\begin{eqnarray}
ds^2 =g_{AB}dx^{A}dx^{B}=g_{00} c^{2} dt^2 + g_{ij} dx^{i} dx^{j},
\label{metric}
\end{eqnarray}
and the square of the four-momentum in this background is
\begin{eqnarray}
p^{A}p_{A} = g_{00}(k^0)^2 + g_{ij} k^{i} k^{j} (1-\alpha {\bf{k}} + 2\alpha^2{\bf{k}}^2)^2~.
\end{eqnarray}
The energy of a particle can be expressed in terms of high energy
momentum as follows  \cite{Vagenas13}
\begin{eqnarray}
{E^2} = {\left( { - {g_{AB}}{\xi ^A}{p^B}}
\right)^2}=-{g_{00}}\left( {{m^2}{c^4}  + {c^2}{p^2}(1 + 2\alpha p)}
\right) ,
\end{eqnarray}
where $\xi^A=(1,0,0,...)$ is the killing vector. Now, the energy of
a particle is $\frac{E}{c}=-g_{AB}\xi^{A} p^{B}$ and the energy in
the gravitational background with metric (\ref{metric}) is given as
$E = - g_{00} c p^0$. Here, we work in the Minkowski spacetime in
which $g_{00} = -1$.

Notice that, this procedure is similar to the modification of the
Peierls-Landau relativistic uncertainty relation which is first
proposed by Amelino-Camelia \emph{et al.} \cite{Camelia066}. To this
end, after simple calculation (neglecting the rest mass), we obtain
\begin{equation}\label{dE}
\frac{d E}{d p} \simeq c\sqrt { - {g_{00}}} \left( {1 + 2\alpha p -
\frac{3}{2}{\alpha ^2}{p^2}}  \right),
\end{equation}
to ${\cal{O}}(\alpha ^2)$. Following the heuristic argument  of
Refs.~\cite{Camelia06,Camelia066,Camelia13}, based on MDR , and
using  $p \simeq \frac{E}{{c\sqrt { - {g_{00}}} }}{\left( {1 -
\frac{{\alpha E}}{{c\sqrt { - {g_{00}}} }}} \right)}$, we have
\begin{eqnarray}
\delta E = \left( c\sqrt { - g_{00}}  + 2 E\alpha  + \frac{7 E^2}{2
c\sqrt { - g_{00}} }\alpha^2+{\cal{O}}(\alpha ^3)\right)\delta p.
\end{eqnarray}
Now, taking $\delta E\simeq E$ and $ E \ge \frac{1}{\delta x}$ which
is suggested by quantum  field theory we find
\begin{eqnarray}
E\delta x \ge c\sqrt { - {g_{00}}} \left( {1 + \frac{{2\alpha
}}{{c\sqrt { - {g_{00}}} \delta x}} - \frac{{7{\alpha
^2}}}{{2{c^2}{g_{00}}{{\left( {\delta x} \right)}^2}}}} \right).
\end{eqnarray}

There are two points should be considered here. First, according to
Eq.~(\ref{momentum}) we keep all terms up to order of $\alpha ^2$
such that considering more generalized form of MDR* will not change
these results. Second, if we consider all natural cut offs such as
minimal length and maximal momentum, not only even powers of energy
but also odd powers of energy should be present
\cite{Sefiedgar10,Sefiedgar11,Sefiedgar12}. Notice that, in the
absence of quantum gravity corrections, i.e., $\alpha=0$, we obtain
the standard dispersion  relation $E^2=m^2c^4+c^2k^2$. Also, in the
following sections we set $\hbar = c = k_B = 1$.

\section{ Reissner-Nordstr\"{o}m (RN) black holes in extra dimensions}\label{sec3}
The RN black hole is a solution of the Einstein equation coupled to
the Maxwell field \cite{Altamirano14,Miao11}. Let us now consider
the RN black hole thermodynamics in universes with large extra
dimensions. There are many scenarios of LED such as Randall-Sundrum
\cite{Randall99} , Arkani-Hamed-Dimopoulos-Dvali
(ADD)\cite{Arkani99} and Dvali-Gabadadze-Porrati \cite{Dvali00}.

In LED scenario, RN metric can be written as follows \cite{Aman06}
\begin{equation}
ds^2 = -F(r) dt^2 +\frac {dr^2}{F(r)} + r^2 d\Omega^2_{D-2},
\end{equation}
where
\begin{eqnarray}
F(r) = 1 - \frac{{2M}}{{{r^{D - 3}}}} + \frac{{{Q^2}}}{{{r^{2(D -
3)}}}},
\end{eqnarray}
and $d\Omega^2_{D-2}$ is the line element on the $(D-2)$-dimensional
unit sphere and the volume of the $(D-2)$-dimensional unit sphere is
given by $\Omega_{D-2} =
\frac{2\pi^{\frac{D-1}{2}}}{\Gamma(\frac{D-1}{2})}$.  The mass and
electric charge of the black hole are given by
\begin{equation}
M = \frac{8\pi G_D }{(D-2) \Omega_{D-2}}m, \qquad Q=
\sqrt{\frac{8\pi G_D}{(D-2)(D-3)}} \:q .
\end{equation}
Here, $G_D$ is gravitational constant in $D$-dimensional spacetime
such that in ADD model is given by
\begin{equation}
G_{D} = \frac{(2\pi)^{D-4}}{\Omega_{D-2}}M_{Pl}^{2-D},
\end{equation}
where $M_{Pl}$ is the $D$-dimensional Planck mass and there is an
effective 4-dimensional Newton constant related to $M_{Pl}$ by
\begin{equation}
M_{Pl}^{2-D}=4\pi G_{4}R^{D-4},
\end{equation}
where $R$ is the size of extra dimensions. It is necessary to note
that in this work, the conventions for definition of the fundamental
Planck scale $M_{Pl}$ are the same as which have been used by ADD.
The location of the outer and inner horizons,  determined by
$F(r)=0$, are given by
\begin{eqnarray}
{{r_\pm } = {{\left( {M\pm\sqrt {{M^2} - {Q^2}} }
\right)}^{\frac{1}{{D - 3}}}}}, \qquad {{M^2} \ge {Q^2}},
\end{eqnarray}
and the horizon area is given by $A_D={\Omega_{D-2}}{r^{D-2}_+}$.
Moreover, the entropy reads $S=\frac {A_D}{4}$.

\section{MDR* and the entropy of RN black holes}\label{sec4}
We are now interested to calculate the microcanonical entropy of RN
black hole. Following heuristic considerations due to Bekenstein,
the minimum increase of the area of a BH absorbing a classical
particle of energy $E$ and size $R$ is given by
\cite{Christodoulou71,Mehdipour05} (After correcting the calibration
factor)
\begin{equation}
(\Delta A)_{min}\geq{4 \ln(2) L_{Pl}^{D-2}E R},
\end{equation}
where $R\sim\delta x\sim r_+$ and $\delta x={\left(
{\frac{{A}}{{\Omega _{D - 2}}}} \right)}^{\frac{1}{{D - 2}}}$. If we
set $(\Delta S)_{min}=\ln2$, then we find
\begin{eqnarray}
\frac{{dS}}{{dA}} \simeq \frac{{{{(\Delta S)}_{min}}}}{{{{(\Delta
A)}_{min}}}} \simeq \frac{{1}}{{4 L_{Pl}^{D - 2}E\delta x}}  =
\frac{1}{{4 L_{Pl}^{D - 2} \Phi(\delta x)}},
\end{eqnarray}
and
\begin{eqnarray}
S_{MDR^*} = \int\limits_{{A_p}}^A {\frac{{dA}}{{4 L_{pl}^{D -
2}\left( {1 + 2\alpha {{\left( {\frac{{A}}{{\Omega _{D - 2}}}}
\right)}^{\frac{1}{{D - 2}}}}  - \frac{{7{\alpha ^2}}}{2}{{\left(
{\frac{{A}}{{\Omega _{D - 2}}}} \right)}^{\frac{2}{{D -
2}}}}}\right)}}}.
\end{eqnarray}

The existence of a minimal length and a maximal momentum leads to
the presence of a minimum event horizon area,  $A_p ={\Omega_{D-2}}
(\delta x)_{min}^{D-2}={\Omega_{D-2}}(\alpha L_{Pl})^{D-2}$
\cite{Pedram11}. After some calculations, the RN black hole entropy
reads
\begin{eqnarray}
D=4 &\rightarrow & {S_4} = \frac{A}{4} - 2\alpha \sqrt \pi  \sqrt A  + \frac{{15}}{2}{\alpha ^2}\pi \ln \left( A \right) + \frac{{88{\pi ^{3/2}}{\alpha ^3}}}{{\sqrt A }} - \frac{{281{\pi ^2}{\alpha ^4}}}{A}+\mbox{const.},\\
D=5 &\rightarrow & {S_5} =\frac{A}{4} - \frac{3}{4}\alpha \sqrt[3]{{2{\pi ^2}}}{A^{2/3}} + \frac{{45}}{8}{\alpha ^2}{\left( {2{\pi ^2}} \right)^{2/3}}\sqrt[3]{A} - 11{\pi ^2}{\alpha ^3}\ln \left( A \right) - \frac{{843}}{{8}}{\alpha ^4}\sqrt[3]{{\frac{{2{\pi ^8}}}{A}}}+\mbox{const.},\hspace{.5cm}\\
D=6 & \rightarrow & {S_6} = \frac{A}{4} - \frac{2}{9} \alpha \sqrt
\pi  {\left( {6A} \right)^{3/4}} + \frac{5}{2} {\alpha ^2}\pi \sqrt
{6A}  - \frac{{22}}{3}{\alpha ^3}\sqrt[4]{{3{\pi
^6}}}\sqrt[4]{{512A}} + \frac{{281 }}{6}{\pi ^2}{\alpha ^4}\ln
\left( A \right)+\mbox{const.}\hspace{.5cm}
\end{eqnarray}

There are many discussions concerning logarithmic corrections to the
entropy area relation \cite{Sefiedgar11,Sefiedgar12,ata1,ata2}. The
logarithmic corrections to black hole have been also obtained in the
tunneling formalism \cite{m1,m2,m3,m4,m5,m6}. The logarithmic
prefactor contains some information about the details of the
underlying quantum gravity proposal. Here, the logarithmic prefactor
is given by $c_0=\frac{{15}}{2} {\alpha ^2}\pi$ for the RN black
hole with double horizons. Thus, we find that this procedure as
mentioned in Ref.~\cite{Hai-Xia07} is not only valid for single
horizon spacetime but also valid for spacetimes with outer and inner
horizons.

Now we easily conclude that the logarithmic prefactor will be
appeared for all number of dimensions. In addition, for positive
values of $\alpha$, the sign of the logarithmic factor is positive
for even number of dimensions but is negative for odd number of
dimensions. This result is the main difference between new form of
MDR with other quantum gravity approaches. In addition, we conclude
that the existence of the logarithmic prefactor is independent of
the dimensionality of the spacetime but depends on the used
statistical ensemble.

\begin{figure}[ht]
\centering
\begin{minipage}[b]{0.45\linewidth}
\includegraphics[width=\linewidth]{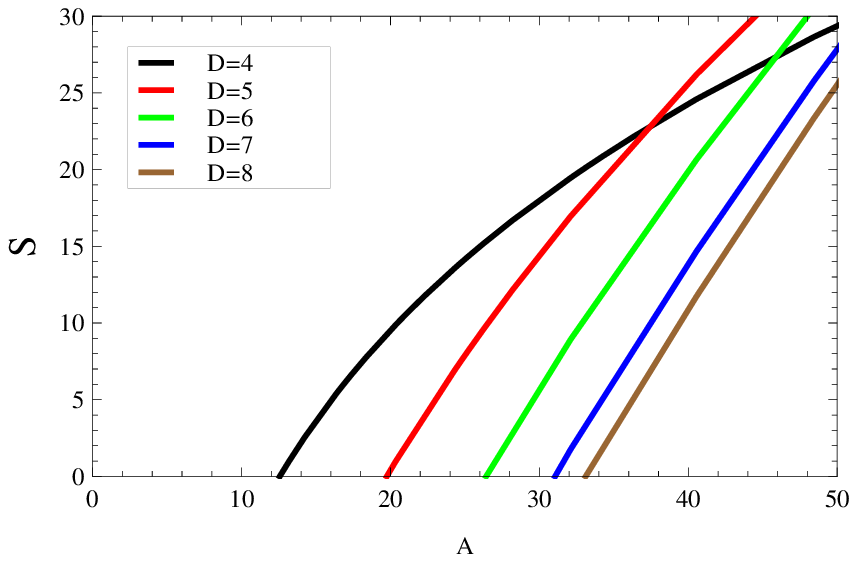}
\caption{\label{fig:s} \small Modified Reissner-Nordstr\"{o}m black
hole's entropy as a function of event horizon area  for different
numbers of spacetime dimensions in the presence of MDR* for
$\alpha=1$.}
\end{minipage}
\quad
\begin{minipage}[b]{0.45\linewidth}
\includegraphics[width=\linewidth]{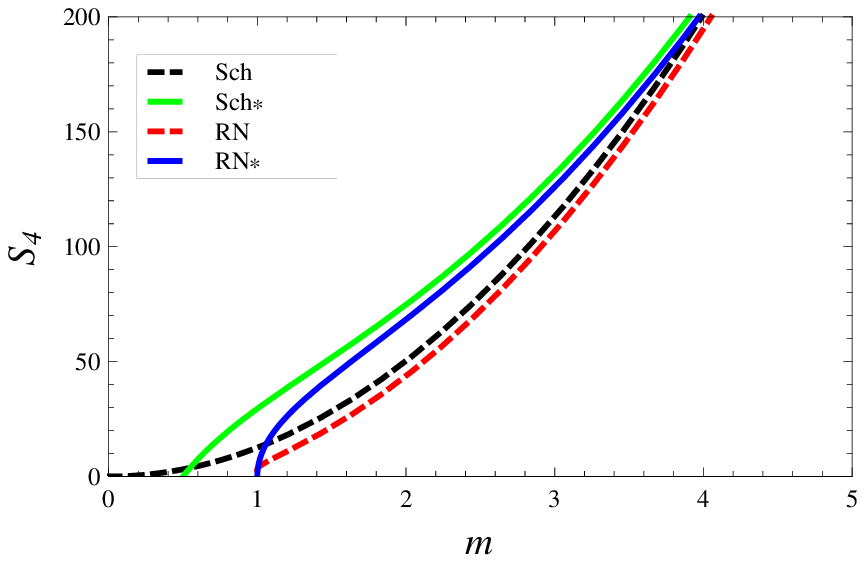}
\caption{\label{fig:s4} \small Entropy of Schwarzschild $(Sch)$,
Reissner Nordstr\"{o}m $(RN)$, modified Schwarzschild  $(Sch*)$, and
modified Reissner Nordstr\"{o}m $(RN*)$ black holes  for $D = 4$.}
\end{minipage}
\end{figure}

Figure \ref{fig:s} shows the relation between the event horizon area
and the entropy of the RN black hole. In scenarios with extra
dimensions, black hole entropy decreases. The classical picture
breaks down since the degrees of freedom of the black hole are
small. In this situation one can use the semiclassical entropy to
measure the validity of the semiclassical approximation. Also black
holes in extra dimensional models have less entropy than black holes
in four dimensions.

Figure \ref{fig:s4} displays the black hole entropy versus its mass
for $4$-dimensional Schwarzschild and Reissner Nordstr\"{o}m black
holes in the presence and absence of MDR*. Therefore, higher
dimensional black hole remnants have less classical features
relative to their four dimensional counterparts, and the mass of the
black hole remnant usually increases with the spacetime dimension
$D$ \cite{ata1,ata2}. It is worth mentioning that in the classical
viewpoint, at the end of the evaporation process, the black hole
contains zero remnant mass, zero final entropy and infinite finial
temperature. However, as we will show, we obtain a nonzero remnant
mass, nonzero final entropy, and finite final temperature.

\section{MDR* and the temperature of RN black holes}\label{sec5}
The Hawking temperature for the spherically symmetric black holes
has been obtained in several ways using MDR
\cite{Camelia06,ata1,Mehdipour08}. According to the first law of the
RN black hole, we have
\begin{eqnarray}
dM = \frac{\kappa }{{8\pi }}dA + \phi dQ = TdS + \phi dQ .
\end{eqnarray}
In more general situations, the entropy of black hole is assumed to
be a function of its area, namely, $S =S(M,Q)$.  The temperature is
expressed as
\begin{eqnarray}
T_{MDR^*}= {\left( {\frac{{\partial M}}{{\partial S}}} \right)_Q} =
\frac{{dA}}{{dS}} \times {\left( {\frac{{dM}}{{dA}}} \right)_Q}  =
\frac{{dA}}{{dS}} \times \frac{\kappa }{{8\pi }} .
\end{eqnarray}
The surface gravity $\kappa(M,Q)$ can be obtained in the usual
manner as \cite{Aman06,Angheben05}
\begin{eqnarray}
\kappa ={1\over 2}|\partial_r {F(r)}|_{r=r_+} .
\end{eqnarray}
Now, using ${T_{BH}} = \frac{{\hbar \kappa }}{{2\pi }}$, the
modified temperature of RN black hole reads
\begin{eqnarray}\label{t}
T_{MDR^*} = \frac{{(D - 3)}}{{4\pi {r_+}}}\left( {1 - \frac{{{\chi _
- }}}{{{\chi _ + }}}} \right)\Phi({r_+}),
\end{eqnarray}
where
\begin{eqnarray}
{\chi _ + } = M + \sqrt {{M^2} - {Q^2}};\quad {\chi _ - } = M -
\sqrt {{M^2} - {Q^2}}.
\end{eqnarray}
This relation shows implicitly that the black hole's temperature
increases with the spacetime's dimension $D$. The higher temperature
leads to faster decay and less classical properties of the black
hole. As a result, both the temperature and the entropy of the RN
black hole receive important corrections such that the temperature
is bounded from above. Such remnants of black holes may be
considered as a candidate for cold dark matter.

\begin{figure}[ht]
\centering
\begin{minipage}[b]{0.45\linewidth}
\includegraphics[width=\linewidth]{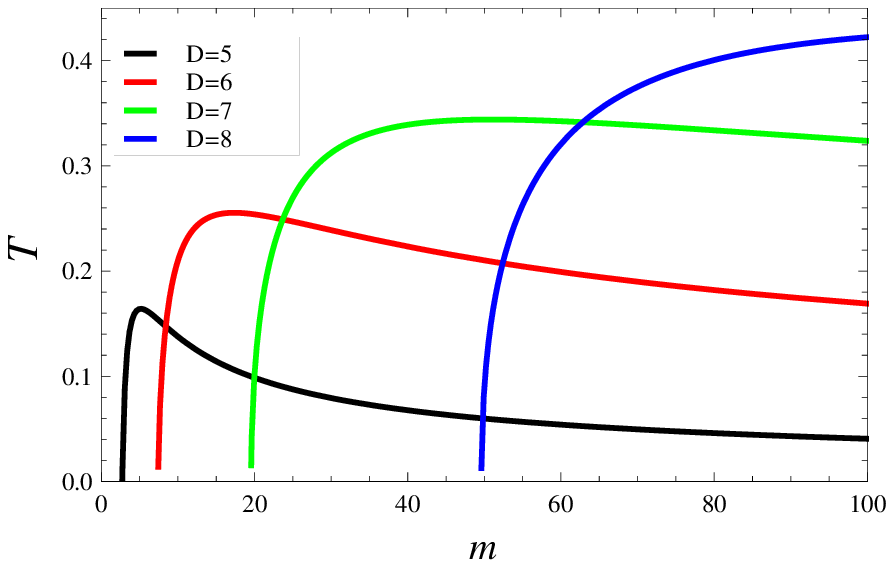}
\caption{\label{fig:t} \small Modified Reissner-Nordstr\"{o}m Black
hole's temperature as a function  of mass for different numbers of
spacetime dimensions in the presence of MDR* for $\alpha=1$ and
$q=1$.}
\end{minipage}
\quad
\begin{minipage}[b]{0.45\linewidth}
\includegraphics[width=\linewidth]{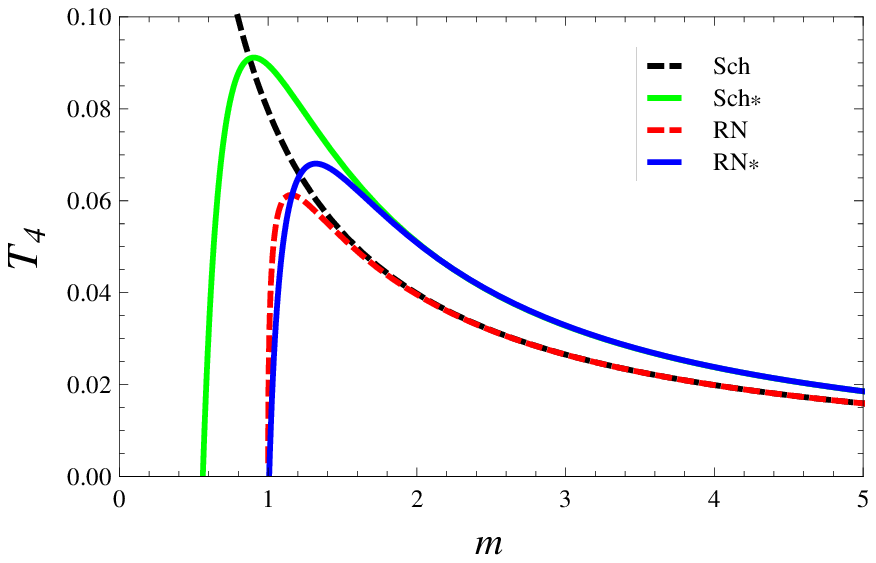}
\caption{\label{fig:t4} \small Temperature of Schwarzschild $(Sch)$,
Reissner Nordstr\"{o}m $(RN)$,  modified Schwarzschild $(Sch*)$ and
modified Reissner Nordstr\"{o}m $(RN*)$ black holes for $D = 4$,
$\alpha=1$ and $q=1$.}
\end{minipage}
\end{figure}

For $D=4$ we have
\begin{eqnarray}
T_4= \frac{{\sqrt {{m^2} - {q^2}} }}{{2\pi {{(m + \sqrt {{m^2} -
{q^2}} )}^2}}} \left( {1 + \frac{{2\alpha }}{{m + \sqrt {{m^2} -
{q^2}} }} - \frac{{7{\alpha ^2}}}{{2{{(m + \sqrt {{m^2} - {q^2}}
)}^2}}}} \right).
\end{eqnarray}
If we set $\alpha=0$, temperature reduces to result that has been
reported in Ref.~\cite{Miao11}.  Also, for $\alpha=0$ and $q=0$, we
obtain the well-known Hawking temperature, i.e., $T_{BH}=\frac{1}{8
\pi m}$. Indeed, our result contains all limiting cases properly.

As Fig.~\ref{fig:t} shows, the black hole radiates until it reaches
to the minimum mass. During this process, its effective temperature
reaches a maximum value. However, as the figure exhibits, when the
radiation stops, the temperature goes to zero. We can say that, when
the radiation reaches to its endpoint and the entropy becomes zero,
and the temperature is in its maximum value, there is a remnant of
black hole. Note that, remnants do not need horizon structure
\cite{Adler01, ata2}.

Figure \ref{fig:t4} shows the comparison between the temperature of
the Schwarzschild black hole, RN black hole and their modified
temperature in the presence of MDR* for $D=4$. It shows that the
temperature of RN black hole remnant is smaller than the temperature
of Schwarzschild black hole remnant. So, the RN black hole remnant
is colder than the Schwarzschild black hole remnant and the natural
cut offs become more effective for small charges. Also, the
temperature of the RN black hole and the modified RN black hole are
distinct.

As another important outcome, according to
\cite{Fazlpour08,Islamzadeh13} the noncommutative Schwarzschild
black hole has features very similar to a commutative RN black hole.
Indeed, it is shown that there is a close connection between charge
and noncommutativity \cite{Fazlpour08,Islamzadeh13}. Here, using
MDR* and comparing evaporation process of the standard and modified
RN black hole, we conclude that there is a nontrivial connection
between charge and the dimensionality of the spacetime near the
Planck scale.
\begin{figure}[ht]
\centering
\begin{minipage}[b]{0.45\linewidth}
\includegraphics[width=\linewidth]{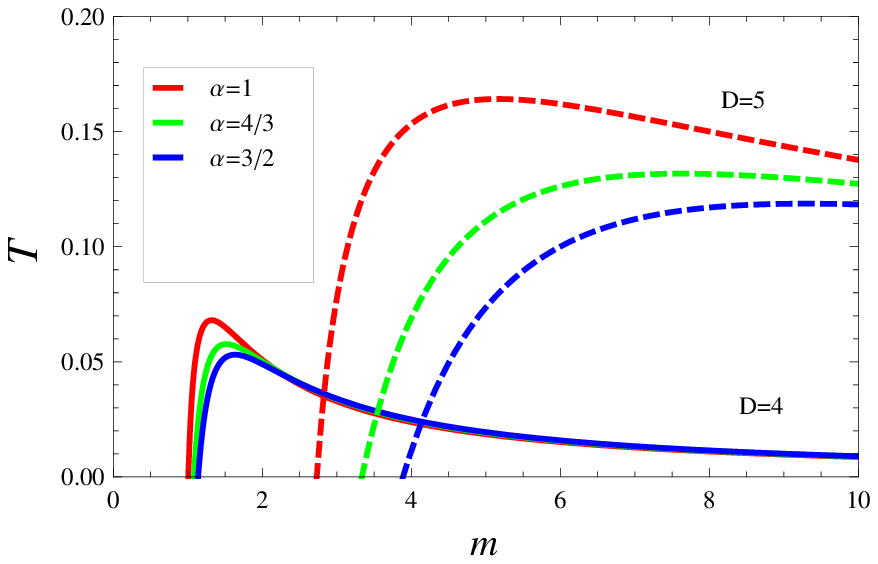}
\caption{\label{fig:t-alpha} \small Modified Reissner Nordstr\"{o}m
black hole's  temperature as a function of mass for different
$\alpha$ ($L_{pl} = 1$).}
\end{minipage}
\quad
\begin{minipage}[b]{0.45\linewidth}
\includegraphics[width=\linewidth]{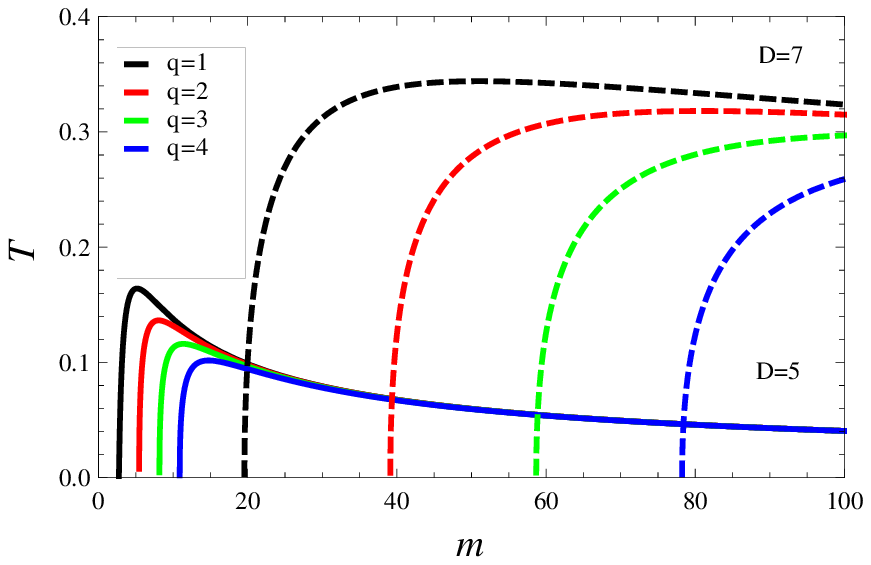}
\caption{\label{fig:t-q}\small Modified Reissner Nordstr\"{o}m black
hole's  temperature as a function of mass for different charges.
($q=1$ and $L_{pl} = 1$).}
\end{minipage}
\end{figure}

Figure \ref{fig:t-alpha} shows that when $\alpha$ increases, the
minimum mass  increases and the maximum temperature decreases.
Figure \ref{fig:t-q} shows that the modified RN black hole
temperature as a function of mass for different values of charge. As
the figure shows, the final state temperature decreases as the black
hole charge increases. We note that black hole evaporates through
the radiation of charged particle-antiparticle pairs until it
reaches a remnant with maximal temperature.

\section{MDR* and the heat capacity of the RN black holes}\label{sec6}
The heat capacity is calculated from the entropy via the relation
\begin{eqnarray}
C=T \left( \frac{\partial S}{\partial T}\right) =\left(
\frac{\partial m}{\partial T}\right) .
\end{eqnarray}
The high energy corrections may prevent the black hole from total
evaporation since the heat capacity vanishes as the temperature
reaches its maximal value. Now, we find the heat capacity of the RN
black hole as a function of its mass. For $D=4$ we have
\begin{eqnarray}
C_{MDR^*} = \frac{{4\sqrt {{m^2} - {q^2}} {{\left( {m + \sqrt {{m^2}
- {q^2}} } \right)}^4}}}{{(28 {\alpha ^2} - 8 \alpha m - 4  {m^2} +
4 {q^2})\sqrt {{m^2} - {q^2}}  - 4 {m^3} - 8 \alpha {m^2} + ( - 7
{\alpha ^2} + 6 {q^2})m + 12 \alpha {q^2}}}.
\end{eqnarray}

\begin{figure}[ht]
\centering
\begin{minipage}[b]{0.45\linewidth}
\includegraphics[width=\linewidth]{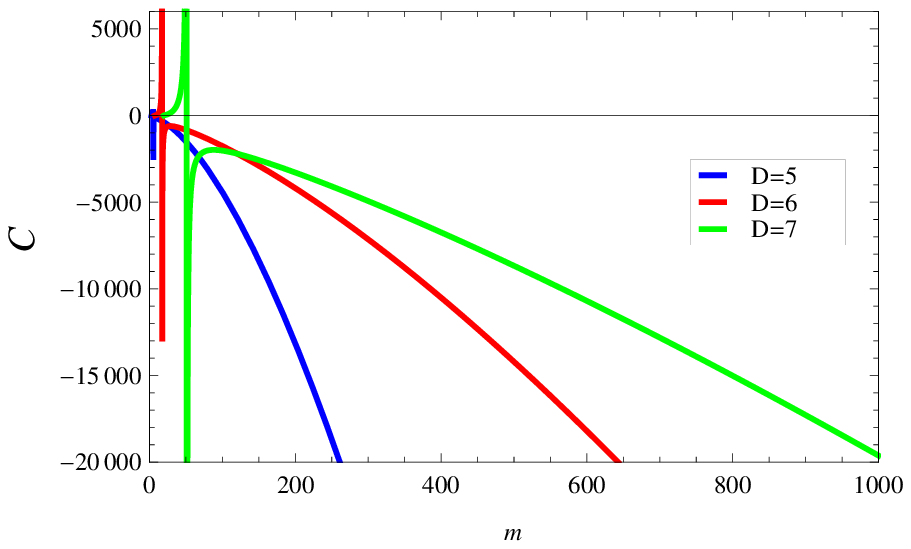}
\caption{\label{fig:c} \small Modified Reissner-Nordstr\"{o}m black
hole's heat capacity as a function of mass for different numbers of
spacetime dimensions in the presence of MDR* for $\alpha=1$ and
$q=1$.}
\end{minipage}
\quad
\begin{minipage}[b]{0.45\linewidth}
\includegraphics[width=\linewidth]{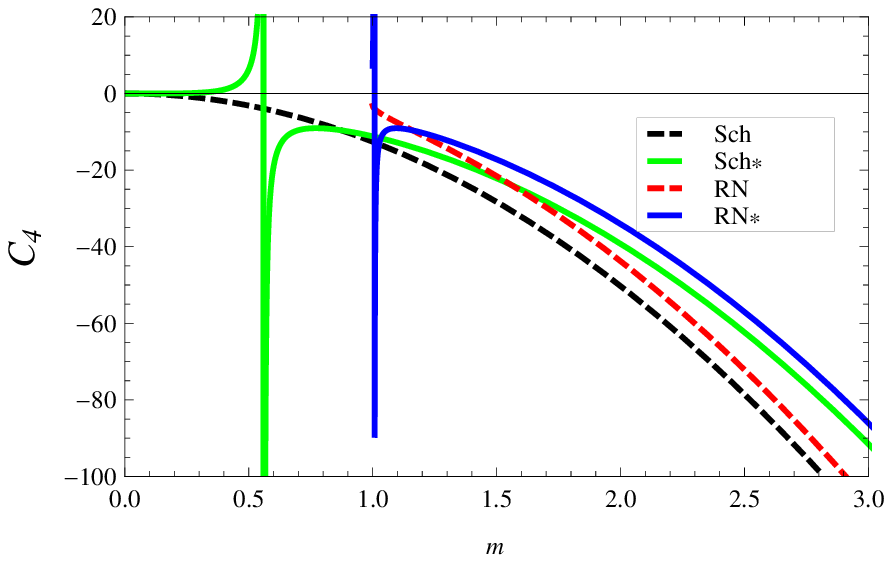}
\caption{\label{fig:c4} \small Heat capacity of Schwarzschild
$(Sch)$,  Reissner Nordstr\"{o}m $(RN)$, modified Schwarzschild
$(Sch*)$ and modified RN $(RN*)$ black holes for $D=4$, $q=1$, and
$\alpha=1$.}
\end{minipage}
\end{figure}

As it can be seen from Fig.~\ref{fig:c}, the negative heat capacity
shows that the thermodynamical system is unstable and tends to
decay. When heat capacity reaches to zero, the system goes to
stability. Indeed, the black hole cannot radiate further and becomes
an inert remnant, possessing only gravitational interactions.

The heat capacity as a function of mass for $D=4$ is represented in
Fig.~\ref{fig:c4}. There is a discontinuity point for the heat
capacity in the modified RN black hole so-called $m_{ext}$ that
leads to a catastrophic evaporation. When the mass of the RN black
hole is above $m_{ext}$, the heat capacity is positive and it tends
to a finite value when mass goes to infinity. It seems this behavior
arises from failure of the standard thermodynamics near the Planck
scale. So, a solution to this problem will also correct the
catastrophic behavior \cite{ata1,ata2,Mehdipour09}.

\section{MDR* and the emission rate of the RN black holes}\label{sec7}
The energy radiated per unit time (the emission rate) can be
calculated using the Stefan-Boltzmann law by assuming that the
energy loss is dominated by photons. In $n_i$-dimensional brane, the
energy radiated by a black body of temperature $T$ and surface area
$A(M,{n_i},D)$ is given by
\cite{Emparan00,Cavaglia00,Cavaglia03,Cavaglia04,Cavaglia033}
\begin{eqnarray}
\frac{{d{E_{n_i}}}}{{dt}} = {\sigma _{n_i}}{A(M,{n_i},D)}{T^{n_i}} .
\end{eqnarray}
We assume that the RN black hole induced area depends on $n_i$, $M$,
and the dimension of the spacetime $D$. So, the RN geometric area
which is induced on the $n_i$-dimensional subspace is
\begin{equation}
A_i(M,n_i,D)=\Omega_{n_i-2}r_c^{n_i-2},
\end{equation}
where $\Omega_{n_i-2}$ is the area of the unit $(n_i-2)$-dimensional
sphere and $r_c=\left(\frac{D-1}{2}\right)^{\frac{1}{D-3}}
\left(\frac{D-1}{D-3}\right)^{1 \over 2}r_+$ is the radius of the
$D$-dimensional  RN black hole of radius $r_+$. Also,
${\sigma_{n_i}}$ is the ${n_i}$-dimensional Stefan-Boltzmann
constant defined as ${\sigma_{n_i}}=\frac{{{\Omega _{{n_i} -
3}}\Gamma ({n_i})\xi ({n_i})}}{{({n_i}- 2){{(2\pi )}^{{n_i}- 1}}}}$.

The thermal emission in the bulk of the brane can be neglected and
the RN black hole is supposed to radiate mainly on the brane
\cite{Emparan00}, i.e.,
$\frac{\frac{{d{E_{4}}}}{{dt}}}{\frac{{d{E_{11}}}}{{dt}}}\simeq 1$.
Thus, the emission rate on the brane is given by
\begin{eqnarray}
{\left( {\frac{{dm}}{{dt}}} \right)_{MDR^*}} \propto  - \lambda
T_{MDR^*}^4,
\end{eqnarray}
where $\lambda= \frac{{{\Omega _1}{\Omega _2}\Gamma (4)\xi
(4)}}{{2{{(2\pi )}^3}}}\left( {\frac{{D - 1}}{{D - 3}}}
\right){\left( {\frac{{D - 1}}{2}} \right)^{\frac{2}{{D -
3}}}}r_+^2$ .

\begin{figure}[ht]
\centering
\begin{minipage}[b]{0.45\linewidth}
\includegraphics[width=\linewidth]{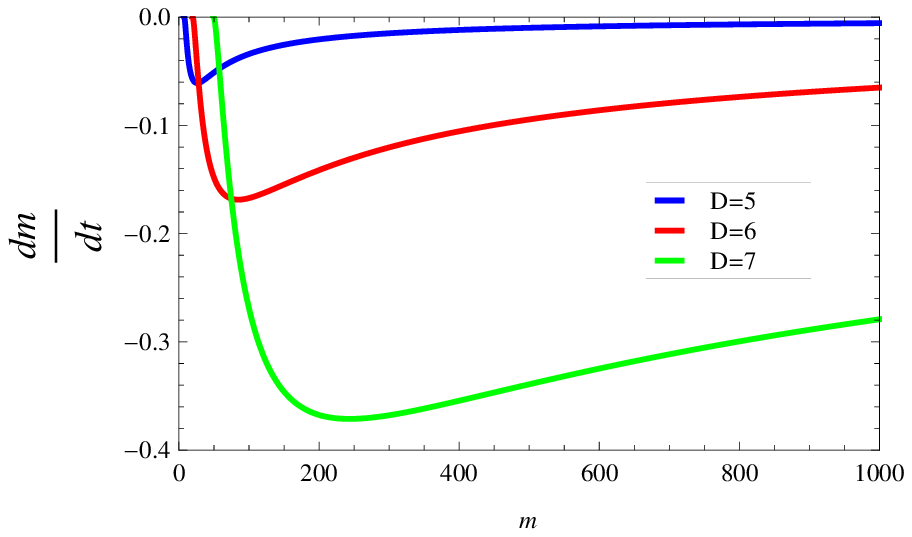}
\caption{\label{fig:dm} \small Emission rate of modified
Reissner-Nordstr\"{o}m black hole as a  function of mass for
different numbers of spacetime dimensions in the presence of MDR*
for $\alpha=1$ and $q=1$.}
\end{minipage}
\quad
\begin{minipage}[b]{0.45\linewidth}
\includegraphics[width=\linewidth]{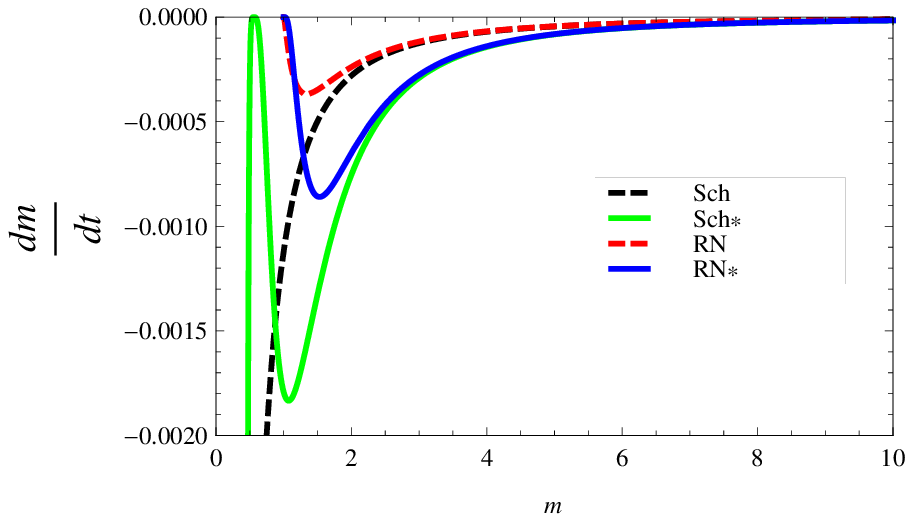}
\caption{\label{fig:dm4} \small Emission rate of Schwarzschild
$(Sch)$, Reissner  Nordstr\"{o}m $(RN)$, modified Schwarzschild
$(Sch*)$ and modified RN $(RN*)$ black holes  for $D=4$, $q=1$, and
$\alpha=1$.}
\end{minipage}
\end{figure}

The emission rate of black hole is shown in Fig.~\ref{fig:dm}. This
means that the emission rate of the RN black hole vanishes when the
black hole reaches its minimal value. In the standard framework, the
emission rate goes to infinity as the mass of the RN black hole
tends to zero. In the MDR* picture, the modified emission rate of
the RN black hole never diverges, and it just goes to zero when the
black hole's mass reaches its minimal value. Also,
Fig.~\ref{fig:dm4} shows the relation between the emission rate of
the Schwarzschild black hole, RN black hole and their modified
temperature in the presence of MDR* for $D = 4$.

\section{Conclusions}\label{sec8}
In this paper, we have studied the effects of a recently proposed
MDR* on the charged black holes. We showed that the presence of a
minimal length and a maximal momentum results in the modification of
RN thermodynamics. Indeed, it leads to faster decay and less
classical behaviors for black holes. We discussed the existence of
the logarithmic prefactor and the relation between the
dimensionality and the entropy. Also, we obtained the temperature,
heat capacity and emission rate of the RN black hole in the presence
of the extra dimensions and compared our result with the standard
formalism. In addition, we discussed the failure of standard
thermodynamics near the Planck energy scale. We showed that in the
modified formalism, the RN black hole has a remnant. The existence
of the remnant has also been predicted  in the context of
noncommutative geometry, Rainbow gravity, GUP and MDR in
\cite{ata1,Ali12,RG2,Camelia13,Fazlpour08,Islamzadeh13}. Thus, it
seems that not only the MDR* but also above approaches of quantum
gravity predict the absence of an effective horizon and the
existence of the remnant for all black holes. This method is valid
for both the single horizon spacetimes and  symmetric spacetimes
with double horizons (outer and inner horizons) and it offers a new
way for studying the entropy corrections of the complicated
spacetimes.

\end{document}